\documentclass{jltp}
\usepackage{graphicx}

\title{Nucleation in a Fermi liquid at negative pressure} 
\author{Fr{\'e}d{\'e}ric Caupin, S{\'e}bastien Balibar and Humphrey J. Maris$^*$}
\address{Laboratoire de Physique Statistique de l'Ecole Normale Sup\'erieure \\
 associ\'e aux Universit\'es Paris 6 et Paris 7 et au CNRS \\
 24 rue Lhomond 75231 Paris Cedex 05, France\\
$^*$Department of Physics, Brown University, Providence, Rhode Island 02912\\
(31 August 2001)}
\runninghead{Fr{\'e}d{\'e}ric Caupin, S{\'e}bastien Balibar and Humphrey J. Maris}{Nucleation in a Fermi liquid at negative pressure}

\newcommand{\mrm}[1]{{\rm{#1}}}
\newcommand{\Eb}{E_\mrm{b}}
\newcommand{\beq}{\begin{equation}}
\newcommand{\eeq}{\end{equation}}
\newcommand{\Rc}{R_\mrm{c}}
\newcommand{\kb}{k_\mrm{B}}
\newcommand{\Pcav}{P_\mrm{cav}}
\newcommand{\Ps}{P_\mrm{s}}

\newcommand{\etal}{\textit{et~al.}}

\newcommand{\vF}{v_\mrm{F}}

\begin{document}

\maketitle

\vspace{-1cm}
\begin{abstract}
Experimental investigation of cavitation in liquid helium~3 has revealed a singular behaviour in the degenerate region at low temperature. As the temperature decreases below 80{\,}mK, the cavitation pressure becomes significantly more negative. To investigate this, we have extrapolated the Fermi parameters in the negative pressure region. This allowed us to calculate the zero sound velocity, which we found to remain finite at the spinodal limit where the first sound velocity vanishes. We discuss the impact on the nucleation of the gas phase in terms of a quantum stiffness of the Fermi liquid. As a result we predict a cavitation pressure which is nearer to the spinodal line than previously thought.

PACS numbers: 71.10.Ay, 67.55.Jd, 64.60.Qb
\end{abstract}

Cavitation has been studied in both isotopes of helium. Using a focused acoustic wave, a small volume of liquid was quenched into the negative pressure region for a short time, and the pressure threshold at which bubble nucleation becomes likely (the cavitation pressure $\Pcav$) was measured. Our group has performed these experiments in the low temperature region, and reported the existence of two cavitation regimes in helium~4\cite{Lambare98,Caupin01}: at high temperature, $\Pcav$ becomes less negative when the temperature increases, and below $600\,\mrm{mK}$, it is temperature independent. This was attributed to a crossover from a thermally activated regime to a quantum regime. At negative pressures, the liquid is metastable and an energy barrier $\Eb$ must be overcome for liquid-gas separation to occur; $\Eb$ decreases as the pressure is more negative and vanishes at the absolute limit of metastability, the so-called spinodal pressure. At high enough temperature, bubble nucleation is triggered by thermal fluctuations: the lower the temperature, the lower the energy barrier that can be overcome, and the more negative the cavitation pressure. Close to the absolute zero, thermal fluctuations have such a small energy that the most favourable process for nucleation is the quantum tunneling through the barrier; the nucleation rate then becomes temperature independent, as does the cavitation pressure. In helium~4, the observed crossover temperature is consistent with the theoretical predictions\cite{Maris95,Guilleumas96}, once adiabatic cooling in the sound wave is taken into account\cite{Lambare98}.

In liquid helium~3, the situation is somewhat different. At high temperature, the cavitation pressure increases with temperature as expected in the thermally activated regime. However, at low temperature $\Pcav$ does not level to the quantum plateau which was predicted below $120\,\mrm{mK}$; on the contrary, below $80\,\mrm{mK}$ it starts to be significantly more negative with decreasing temperature\cite{Caupin00,Caupin01}. Possible experimental artifacts (thermal gradients\cite{Caupin01}, thermal effect of the sound wave\cite{Balibar98}) were ruled out and the search for a theoretical interpretation began. The viscosity of liquid helium~3 diverges as $1/T^2$ in the Fermi liquid region, that is below $100\,\mrm{mK}$~\cite{Wheatley75}. Dissipation effects were considered by Jezek~{\etal}\cite{Jezek99}: they concluded that the crossover temperature could be lowered to $50\,\mrm{mK}$ with a dissipation parameter corresponding to the measured viscosity at $100\,\mrm{mK}$ under saturated vapor pressure; however, they note that by using the macroscopic viscosity coefficient, overestimating the dissipation effects is likely. Furthermore, lowering the crossover temperature is not sufficient to explain the sharp variation of the cavitation pressure, and we have tried to find another interpretation.

In Fermi liquids, there is a particular mode of sound propagation called zero sound; it was first predicted by Landau\cite{Landau57} and later demonstrated ex\-per\-i\-men\-tal\-ly\cite{Wilks}. In the ballistic regime, when the lifetime $\tau$ of the quasiparticles becomes longer than the period $T_\mrm{sound}=1/f$ of the sound wave, the sound can still propagate because of the interaction between quasiparticles. It was first suggested by Balibar~{\etal}\cite{Balibar98} that at the spinodal, whereas the first sound velocity $c_1$ vanishes, the zero sound velocity $c_0$ could remain finite, and that the liquid would then present a finite stiffness for the high-frequency motions; they explained that this would lower the crossover temperature. In this communication, we present an estimate of the zero sound velocity at the spinodal and we consider its effect on both the quantum and the thermally activated nucleation regime.

The zero sound velocity and the corresponding distortion of the Fermi surface $\nu(\theta,\phi)$ are calculated by solving the following equation\cite{Wilks,BaymPethick}:
\beq
(s-\cos\theta) \,\nu(\theta,\phi)=\cos\theta \int F(\chi)\, \nu(\theta',\phi')\,\frac{\mathrm{d} \Omega'}{4 \pi}\; ,
\label{eq:son0}
\eeq
where $s=c_0/\vF$ ($\vF$ being the Fermi velocity), $\chi$ is the angle between the momentum of two quasiparticles and $\mathrm{d} \Omega^\prime$ is the volume element. $F$ is a function describing the quasiparticles interaction; it is usually replaced by the first terms of its expansion along the base of the Legendre polynomials. The coefficients $F_l$ of this expansion are the Fermi parameters. The first $F_l$ are related to measurable quantities and are thus known experimentally. To estimate the velocity of zero sound at negative pressures we proceed as follows: (1) we first estimate $c_1$ using an extrapolation of measured data at positive pressures using the method proposed by Maris (see Appendix of Ref.~\onlinecite{Caupin01}); (2) Greywall has measured the effective mass $m^\ast$ for positive pressures\cite{Greywall83,Greywall86}. His data can be fit well by the equation
\beq
\frac{m}{m^\ast}=a^2\left[ 1-\frac{\rho}{\rho_\mrm{c}} \right]^2\; ,
\eeq
with $a=1.018$ and $\rho_\mrm{c}=198.6\,\mrm{kg\,m^{-3}}$; the form of this fit was proposed by Stringari\cite{Stringari84} on the basis of a microscopic model. We then use this equation to estimate $m^\ast$ at negative pressure. This then gives $F_1$ through the relation $F_1=3 [(m^\ast/m)-1]$; (3) $F_0$ is then extrapolated using
\beq
F_0 = 3 \left( \frac{m}{m^\ast} \frac{c_1}{\vF} \right)^2 -1\; .
\eeq
(4) Finally, $F_2$ is obtained by means of a simple polynomial extrapolation of the data compiled in Ref.~\onlinecite{Halperin}. It is interesting to note that, at the spinodal, $c_1=0$ implies $F_0=-1$: the classical mechanical instability at the spinodal coincides with a fermionic instability occurring when $F_l=-(2l+1)$~\cite{Pomeranchuk58}.
\begin{figure}
\centerline{\includegraphics[height=4.9cm]{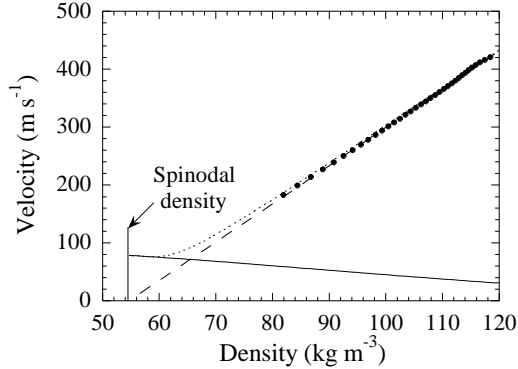}}
\caption{Characteristic velocities in degenerate helium~3: Fermi (solid line), first sound (dashed line) and zero sound (dotted line) velocities. Full circles are experimental data for the first sound velocity taken from Ref.~\protect\onlinecite{Halperin}.}
\label{fig:vitesses}
\end{figure}

By numerically solving Eq.~\ref{eq:son0} with the two first Fermi parameters\cite{theseCaupin01}, we find the zero sound velocity: the result is displayed on Fig.~\ref{fig:vitesses}; taking $F_2$ into account affects our results only slightly (by less than $2.5\,\%$)\cite{theseCaupin01}. This calculation shows that whereas the first and zero sound velocities are close at positive pressure, they depart from each other at negative pressure; we find that at the spinodal, whereas the first sound velocity vanishes, the zero sound velocity becomes equal to the Fermi velocity $\vF\simeq 80\,\mrm{m\,s^{-1}}$. Indeed, for Eq.~\ref{eq:son0} to have a real solution, that is to observe zero sound modes which are not strongly damped, one must have $c_0>\vF$~\cite{BaymPethick}: the departure between $c_0$ and $c_1$ is thus a strong physical effect, independent of the chosen extrapolations.

We now turn to the crossover frequency from first to zero sound at a given temperature $f_{0-1}(T)$ (or equivalently the crossover temperature at a given frequency): this is obtained by solving $2\pi f_{0-1}(T) \tau(T)=1$. The lifetime $\tau$ of the quasiparticles is related to the viscosity $\eta$, the density and the effective mass\cite{Wilks}:
\beq
\tau= \frac{5\eta}{\rho \vF} \frac{m}{m^\ast}\; .
\eeq
In the degenerate Fermi liquid region, $\eta T^2$ is constant at a given density and this constant decreases with density\cite{Wheatley75,Wilks}. At $100\,\mrm{mK}$, this gives a crossover frequency $f_{0-1}$ less than $1.5\,\mrm{GHz}$ at saturated vapor pressure and a polynomial extrapolation of $\eta T^2$ even gives $f_{0-1}<1\,\mrm{GHz}$ near the spinodal; this is to be compared with the thermal frequency $\kb T/h\simeq 2\,\mrm{GHz}$ at $100\,\mrm{mK}$, and to the tunneling frequency in the quantum cavitation regime, $f_\mrm{Q}\simeq10\,\mrm{GHz}$~\cite{Maris95}.

Let us now consider the effect of zero sound on nucleation: density fluctuations propagate in the zero sound mode with a higher phase velocity than in the first sound mode, so that nucleation rates driven by the faster fluctuations should decrease. In the quantum cavitation regime, the nucleation rate is related to the action calculated for the time that the system takes to pass under the barrier. The quantum regime previously predicted\cite{Maris95,Guilleumas96} is a high frequency process which must in fact be described by zero sound, which increases the action and decreases the corresponding nucleation rate. As for the thermally activated cavitation, we also think that the cavitation rate is reduced. The critical nuclei for nucleation were calculated to be spheres of radius $\Rc\simeq1\,\mrm{nm}$; at low temperature, this becomes small compared to the mean free path $\bar{l}$ of the quasiparticles: close to the spinodal, $\bar{l}=\vF \tau\simeq 11\,\mrm{nm}$ at $100\,\mrm{mK}$ and $45\,\mrm{nm}$ at $50\,\mrm{mK}$. The thermal fluctuations leading to these critical nuclei involve a superposition of collective modes with different wavelengths, centered on the wavelength $\Rc$; consequently, most of these fluctuations must be described by zero sound. We believe that the present approach based on Landau's theory is valid at this scale because the Fermi length $l_\mrm{F}=h/(m^\ast \vF)\simeq 0.14\,\mrm{nm}$ at the spinodal remains small compared to $\Rc$, so that a local Fermi surface can be defined; this surface undergoes a non-spherically symmetric deformation because of the radial mass transfer which occurs during the evolution from the homogeneous liquid to the critical profile, with a density hollow at the center. The energy of one phonon of wavelength $\Rc$ is $h c_0/\Rc$, instead of $h c_1/\Rc$: the ratio, $c_1/c_0$, is less than $0.35$ for $\Pcav-\Ps<10\,\mrm{mbar}$, which is the pressure range calculated by previous theories at low temperature in the conditions of the experiment. There will therefore be fewer fluctuations leading to a critical nucleus than expected before. We thus believe that in both regimes, the nucleation rate is decreased, and that cavitation only occurs if all the critical fluctuations are described by first sound.
\begin{figure}
\centerline{\includegraphics[height=4.9cm]{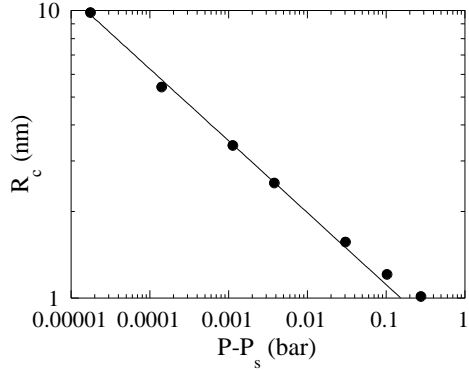}}
\caption{Critical radius as a function of the pressure distance to the spinodal $P-\Ps$. The full circles corresponds to our numerical calculations and the solid line is a fit by a $-1/4$ power law.}
\label{fig:Rc}
\end{figure}

We can evaluate where this different regime should take place: it requires $\Rc > \bar{l}$. This condition is fulfilled at pressure close enough to the spinodal pressure where $\Rc$ diverges. This divergence was predicted by Lifshitz and Kagan\cite{Lifshitz72}, who made an analytical calculation in a generic case. Following Xiong and Maris\cite{Xiong89}, we have calculated numerically the critical density profiles with Maris' equation of state for liquid helium; this equation of state has a singular behavior which Maris justifies with a renormalization calculation\cite{Maris91}. Our calculations give the critical radius $\Rc$ in the region very close to the spinodal. Fig.~\ref{fig:Rc} shows that $\Rc$ diverges following a $-1/4$ power law when the pressure approaches the spinodal. This shows that $\Rc=10\,\mrm{nm}$ when the pressure is at only $2\times10^{-2}\,\mrm{mbar}$ from the spinodal pressure. In the degenerate region, cavitation therefore occurs at pressures closer to the spinodal by three orders of magnitude than what was estimated before.

For these large critical radii, we think that the quantum fluctuations leading to tunneling will have low enough frequency to be described with first sound. But as the crossover temperature from quantum to thermal cavitation decreases when the pressure approaches the spinodal pressure\cite{Xiong89}, it is not clear if the quantum regime still exists and at which temperature it is reached. This point requires further investigation.

In conclusion, we have extrapolated the first Fermi parameters of liquid helium~3 into the negative pressure region. This allowed us to calculate the zero sound velocity. We find that this velocity is non vanishing at the spinodal, contrary to the first sound velocity. This results in a quantum stiffness of the liquid in the degenerate Fermi liquid region. On these grounds, we find that the pressure required for cavitation to occur must be much closer to the spinodal than previously predicted, so that the critical fluctuations remain in the first sound regime. Another consequence is that the crossover from thermally activated cavitation to quantum cavitation is shifted to lower temperatures than predicted by previous theories, if not suppressed. This picture is consistent with the experimental measurements.


\begin{thebibliography}{99}

\bibitem{Lambare98} H.~Lambar\'e, P.~Roche, S.~Balibar, H.~J.~Maris, O.~A.~Andreeva, C.~Guthmann, K.~O.~Keshishev, and E.~Rolley, {\em Eur. Phys. J.} {\bf B2}, 381 (1998).

\bibitem{Caupin01} F.~Caupin and S.~Balibar, {\em Phys. Rev.} {\bf B} {\bf 64}, 064507 (2001).

\bibitem{Maris95} H.~J.~Maris, {\em J. Low Temp. Phys.} {\bf 98}, 403 (1995).

\bibitem{Guilleumas96} M.~Guilleumas, M.~Barranco, D.~M.~Jezek, R.~J.~Lombard, and M.~Pi, {\em Phys. Rev.} {\bf B54}, 16135 (1996).

\bibitem{Caupin00} F.~Caupin and S.~Balibar, {\em Physica} {\bf B284-288}, 212 (2000).

\bibitem{Balibar98} S.~Balibar, F.~Caupin, P.~Roche, and H.~J.~Maris, {\em J. Low Temp. Phys.} {\bf 113}, 459 (1998).

\bibitem{Wheatley75} J. C. Wheatley, {\em Rev. Mod. Phys.}, \textbf{47}, 415 (1975).

\bibitem{Jezek99} D.~M.~Jezek, M.~Pi, and M.~Barranco, {\em Phys. Rev.} {\bf B60}, 3048 (1999).

\bibitem{Landau57} L.~D. Landau, {\em Zh. Eksper. Teor. Fiz.} \textbf{32} 59 (1957) [{\em Soviet Physics JETP} \textbf{5}, 101 (1957)].

\bibitem{Wilks} J.~Wilks, {\em The properties of liquid and solid helium} (Clarendon Press, Oxford, 1967), Chap.~18.

\bibitem{BaymPethick} G. Baym and C. Pethick, {\em Landau Fermi-liquid theory} (J.~Wiley and Sons, New York, 1991), Chap.~1.

\bibitem{Greywall83} D.~S.~Greywall, {\em Phys. Rev.} {\bf B27}, 2747 (1983).

\bibitem{Greywall86} D.~S.~Greywall, {\em Phys. Rev.} {\bf B33}, 7520 (1986).

\bibitem{Stringari84} S. Stringari, {\em Phys. Lett.} {\bf A106}, 267 (1984).

\bibitem{Halperin} W. P. Halperin and E. Varoquaux, in~: {\em Helium~3}, edited by W. P. Halperin and L. P. Pitaevskii (North Holland, Amsterdam, 1990).

\bibitem{Pomeranchuk58} I. Y. Pomeranchuk, {\em Zh. Eksper. Teor.~Fiz.} \textbf{35}, 524 (1958) [{\em Soviet Physics JETP} \textbf{7}, 361 (1959)].

\bibitem{theseCaupin01} F.~Caupin, Ph.~D. thesis, Universit\'e Paris~6, 2001.

\bibitem{Lifshitz72} I.~M.~Lifshitz and Y.~Kagan, {\em Zh. Eksp. Teor. Fiz.} {\bf 62}, 385 (1972) [{\em Sov. Phys. JETP} {\bf 35}, 206 (1972)].

\bibitem{Xiong89} Q. Xiong and H. J. Maris, {\em J. Low Temp. Phys.} \textbf{77}, 347 (1989).

\bibitem{Maris91} H. J. Maris, {\em Phys.~Rev.~Lett.} \textbf{66}, 45 (1991).

\end{thebibliography}
\end{document}